\documentclass[superscriptaddress,aps,pra,twocolumn,amssymb,amsfonts,showpacs,letterpaper]{revtex4}
\usepackage{amsmath}

\input{Qcircuit}

\newcommand{\measureCl}[1]{*+[F-:<.9em>]{#1} \cw}
\usepackage{color}
\usepackage{graphics,graphicx}

\newcommand{\shortqph}[1]{}

\providecommand{\ignore}[1]{}

\newcommand{\mCo}[1]{\textcolor{blue}{}}

\newcommand{\mnote}[1]{\textcolor{blue}{[Manny: #1]}}
\renewcommand{\mnote}[1]{}

\def\openone{\leavevmode\hbox{\small1\kern-3.8pt\normalsize1}}
\def\RR{{\rm I\kern-.2emR}}

\def\openone{\leavevmode\hbox{\small1\kern-3.8pt\normalsize1}}
\def\RR{{\rm I\kern-.2emR}}

\def\cn{{\cal N}}
\def\ch{{\cal H}}
\def\cs{{\cal S}}

\def\cO{{\cal O}}

\def\cz{{\cal Z}}
\def\co{{\cal O}}

\providecommand{\ignore}[1]{}

\newcommand{\sket}[1]{| #1 \rangle}
\newcommand{\sbra}[1]{\langle #1 |}

\newcommand{\inner}[2]{ \langle #1 | #2 \rangle}

\newcommand{\bitem}{\begin{itemize}}
\newcommand{\eitem}{\end{itemize}}
\newcommand{\benum}{\begin{enumerate}}
\newcommand{\eenum}{\end{enumerate}}
\newcommand{\beq}{\begin{equation}}
\newcommand{\eeq}{\end{equation}}
\newcommand{\beqa}{\begin{eqnarray}}
\newcommand{\eeqa}{\end{eqnarray}}
\newtheorem{definition}{Definition}

\newtheorem{proposition}{Proposition}

\newcommand{\bproof}{\begin{proof}}
\newcommand{\eproof}{\end{proof}}
\newcommand{\bprop}{\begin{proposition}}

\newcommand{\bdef}{\begin{definition}}

\def\one{{\mathchoice {1\mskip-4mu {\text {\rm l}}} {
1\mskip-4mu {\text {\rm l}}} {1\mskip-4.5mu {\text {\rm l}}} {1\mskip-5mu
{\text {\rm l}}}}}

\def\({\left(}
\def\){\right)}

\begin{document}

\title{Quantum Simulations of Classical Annealing Processes}

\author{R. D. Somma}
\email{rsomma@perimeterinstitute.ca}

\affiliation{Perimeter Institute for Theoretical Physics, Waterloo, ON N2L 2Y5, Canada }

\author{S. Boixo} \affiliation{CCS-3 (Information Sciences), Los Alamos National Laboratory, Los
  Alamos, NM 87545, USA} \affiliation{Department of Physics and
  Astronomy, University of New Mexico, Albuquerque, NM 87131, USA}
%\email{boixo@unm.edu}

\author{H. Barnum} \affiliation{CCS-3 (Information Sciences), Los Alamos National Laboratory, Los
  Alamos, NM 87545, USA}

%\email{barnum@lanl.gov}

\author{E. Knill} \affiliation{National Institute of Standards and Technology, Boulder, Colorado  80305, USA}

\date{\today}
\begin{abstract}
  We describe a quantum algorithm that solves combinatorial
  optimization problems by quantum simulation of a classical simulated
  annealing process. Our algorithm exploits quantum walks and the
  quantum Zeno effect induced by evolution randomization. It requires
  order $1/\sqrt{\delta}$ steps to find an optimal solution with
  bounded error probability, where $\delta$ is the minimum spectral
  gap of the stochastic matrices used in the classical annealing
  process. This is a quadratic improvement over the order $1/\delta$
  steps required by the latter.
\end{abstract}
\pacs{03.67.Ac, 87.10.Rt, 87.55.de}

\maketitle

%%%%%%%%%%%%%%%%%%%%%%%%%%%%%%%%%%%%%%%%%%%%%%%%%%%%%%%%%%%%%%%%%%%%%%%

%\section{Introduction}
Combinatorial optimization problems (COPs) are important in almost
every branch of science, from computer science to statistical physics
and computational biology~\cite{CCP98}. Each instance of a COP
requires that we minimize some objective function over a search space
consisting of $d$ \emph{configurations}. The search space may have additional
structure, such as that provided by a graph, to give a notion of
locality. Because $d$ is typically exponential in the size of the
problem instance, finding a solution by exhaustive search is hard in
general. One can exploit the notion of locality to find solutions more
quickly, but the presence of many nonoptimal local minima often
prevents efficient convergence to a solution. Therefore, more
efficient optimization strategies are desirable.

A well known and often used general strategy for solving COPs is
simulated annealing (SA)~\cite{KGV83}. SA imitates the process
undergone by a metal that is heated to a high temperature and then
cooled slowly enough for thermal excitations to prevent it from
getting stuck in local minima, so that it ends up in one of its
lowest-energy configurations. In SA, the objective function $E$ of the
COP plays the role of the energy, so the lowest energy configuration
is the optimum. The annealing process can be simulated with a variety
of techniques. Here, we focus on discrete Markov chain Monte-Carlo
(MCMC) as used, for example, in statistical physics~\cite{NB99}. MCMC
generates a stochastic sequence of configurations via a Markov process
that, in the case of SA, converges to the Gibbs distribution at a low
final temperature. More specifically, the annealing process is
determined by a choice of an {\em annealing schedule} consisting of a
finite increasing sequence of {\em inverse temperatures} $\beta_1 <
\beta_2< \ldots <\beta_P$, and by an associated sequence of {\em
transition rules} $\{ M_1,\cdots,M_P\}$ consisting of stochastic
matrices acting on configurations. When the structure of the problem
can be exploited by a good choice of transition rules, the MCMC
algorithm can outperform exhaustive search.

One way to characterize the implementation complexity of SA based on
MCMC is to count the number of times that the transition rules must be
applied before converging to the desired final distribution within an
acceptable error. For simplicity, we consider regular annealing
schedules with $\beta_k = (k-1)\Delta\beta$ and choose $\Delta\beta =
\co(\delta/E_M)$, where $\delta$ is the minimum spectral gap of the
matrices $M_k$ and $E_M = \max_\sigma |E[\sigma]|$. We assume that $E$
has been shifted so that $E\geq 0$. Let $\gamma$ be the spectral gap
of $E$, defined as the difference between the two smallest values in
the range of $E$. By adapting arguments from Ref.~\cite{St05} to the
discrete-time setting it can be shown that if $\beta_P =P \Delta
\beta=\co( \gamma^{-1} \log(d/\epsilon^2))$, then the probability that
SA does not return an optimal configuration is no greater than
$\epsilon$. Thus, for a success probability greater than $1-\epsilon$,
the implementation complexity of SA is given by $\cn_{SA} =P= \co\(
\frac{E_M}{\gamma} \log{(d/\epsilon^2)} / \delta \)$.

Ideally, $\cn_{SA}$ is small compared to the size $d$ of the configuration
space. Since problem instance sizes are typically polylogarithmic in
$d$, $\cn_{SA} = \cO(\textrm{polylog}(d))$ is considered efficient.
Efficient $\cn_{SA}$ is obtained, for example, when computing physical
properties of the $N$-spin ferromagnetic Ising model in an homogeneous external field~\cite{JS93}. However, 
inefficient $\cn_{SA}$ is obtained if the external field is
random~\cite{Ba82}, making the problem intractable due to gaps
$\delta$ that are exponentially small in $N$. The dependence of the
complexity of MCMC on $\delta^{-1}$ is characteristic of Markov
processes and may be unavoidable~\cite{Al81}. Thus, finding new
methods with better scaling in $\delta$ is very desirable.

Quantum mechanics provides new resources with which to attack
optimization problems~\cite{VARIOUS1,SBO07}. Quantum computers (QCs)
can theoretically solve some problems, including integer number
factorization and unstructured search, more efficiently than
classical computers~\cite{VARIOUS2}. Still, whether a QC could solve
all COPs more efficiently than is possible with classical computers is an
open question. In this Letter we show that QCs can speed up the
simulation of classical annealing processes. We present a method for
transforming instances of MCMC-based SA into a quantum simulated
annealing (QSA) algorithm for which the number of times, $\cn_{QSA}$,
that the transition rules are used is
$\co((E_M/\gamma)^2\log^2(d/\epsilon)\log d/(\epsilon\sqrt{\delta}))$,
a quadratic improvement as a function of $\delta^{-1}$. This
improvement is most significant for hard instances where $\delta \ll
1$. The dependence on $1/\epsilon$ can be improved to
$\textrm{polylog}(1/\epsilon)$. QSA is based on ideas and techniques
from quantum walks~\cite{SZWALKS} and the quantum Zeno effect, where
the latter can be implemented by phase estimation or by randomization
of an evolution period.

This paper is organized as follows. First we describe a
``quantization'' of a reversible, ergodic Markov chain in terms of a
bipartite quantum walk. This is a similarity-transformed version of
the quantum walk used in Refs.~\cite{SZWALKS} to obtain quantum
speedups in search problems. The quantum walk is a unitary operator
acting on the state space obtained by superposition from the
configurations of the COP. We then explain how to transform an
instance of SA by adapting the annealing schedule and applying the
Markov chain quantization. Finally, we analyze the complexity of QSA
to determine the speedup over SA.

\noindent
{\bf Quantum Walks and Markov Chains.} Discrete-time quantum
walks were introduced as the quantum analogues of classical random
walks~\cite{QWALKS}. We focus on the bipartite quantum walks defined
in Refs.~\cite{SZWALKS}.
% where they are used to obtain quantum
%speedups in search problems.
% Such quantum walks,  
% which we describe below, can also be derived from 
% Ref.~\cite{Am}.

Consider a $d$-configuration classical system $\cs$ with energies $E[\sigma]$
for configurations $\sigma$. Denote the space of ground configurations (minimizers of
$E$) by $\mathbb{S}_0$. Consider an ergodic, reversible Markov
process on $\cs$ with transition probabilities $p(\sigma'|\sigma) =
m_{\sigma\sigma'}$ and stationary distribution $\pi^\sigma$.
Reversibility is equivalent to the detailed balance condition
$\pi^\sigma m_{\sigma \sigma'}= \pi^{\sigma '} m_{\sigma ' \sigma}$.
Let $\ch$ be the quantum state space spanned by orthonormal states
$\ket{\sigma}$ for configurations $\sigma$ of $\cs$. In SA,
$\pi^{\sigma}=e^{-\beta E[\sigma]}/\cz$ with $\cz=\sum_\sigma
e^{-\beta E[\sigma]}$ is the Gibbs distribution at some inverse
temperature $\beta$. We assume not only that we have a classical
algorithm to efficiently sample from the distribution
$m_{\sigma\sigma'}$ given $\sigma$, but also that we have an efficient
quantum algorithm that computes the transformation defined by
$\sket{\sigma}\sket{\mathfrak{0}}\mapsto\sket{\sigma}\sum_{\sigma'}\sqrt{m_{\sigma\sigma'}}\sket{\sigma'}$,
with $\sket{\mathfrak{0}}$ an efficiently preparable state of $\ch$ (e.g. a
computational basis state).
Note that this stronger condition is usually satisfied, because for
given $\sigma$, $m_{\sigma\sigma'}$ is non-zero for only polynomially
many $\sigma'$, and the non-zero $m_{\sigma\sigma'}$ can be computed
efficiently by a classical algorithm. 

The bipartite quantum walk is defined on the tensor product
$\ch_A\otimes \ch_B$ of two copies of $\ch$. Following \cite{SZWALKS},
we define isometries $X$ and $Y$ that map states of $\ch$ to states of
$\ch_A \otimes \ch_B$ by
\begin{align}
X \sket{\sigma} &= \sket{\sigma} \sum_{\sigma'} \sqrt{m_{\sigma\sigma'}} \sket{\sigma'}, \\
Y \sket{\sigma'} &= \sum_{\sigma} \sqrt{m_{\sigma'\sigma}} \sket{\sigma}\sket{\sigma'} \;.
\end{align}
Let $D_\pi$ be the diagonal matrix with entries
$\pi^{\sigma}$ on the diagonal. Let $M$ be the matrix with entries
$M_{\sigma'\sigma} = m_{\sigma\sigma'}$. From the detailed balance
condition, $X^\dagger Y = D_\pi^{1/2} M D_\pi^{-1/2}$ is symmetric.
It follows that $X^\dagger Y$ and $M$ have the same eigenvalues
$\lambda_0=1>\lambda_1\geq\ldots\geq\lambda_{d-1}\geq 0$. Let
$\sket{\phi_j}$ be the $\lambda_j$ eigenstate of $X^\dagger Y$. Then
$\sket{\phi_0} = \sum_\sigma\sqrt{\pi^{\sigma}}\sket{\sigma}$, which
upon measurement in the basis $\ket{\sigma}$ has the same probability
distribution as the stationary distribution of the Markov process.

Define unitary operators $U_X$ and $U_Y$ by
\begin{align}
\label{uxy}
U_X \sket{\sigma}\sket{\mathfrak{0}} \equiv X\sket{ \sigma},\quad
U_Y \sket{\mathfrak{0}}\sket{\sigma} \equiv Y \sket{\sigma} \;,
\end{align}
with arbitrary action on other states. Let $P_1$ and $P_2$ be the
projectors onto the subspaces spanned by
$\{\sket{\sigma}\sket{\mathfrak{0}}\}_\sigma$ and $\{U_X^\dagger
U^{\;}_Y \sket{ \mathfrak{0}}\sket{\sigma}\}_\sigma$, respectively.
The reflection operators through $P_i$ are defined by $R_i =
2P_i-\one$. A step of the bipartite quantum walk $W$ based on $M$ is
given by $W=R_2 R_1$. This walk is related to the one used in
Ref.~\cite{SZWALKS} by a unitary, but $\pi^{\sigma}$-dependent,
similarity transformation, which helps avoid amplitude leakage when
$W$ changes in QSA.

The spectrum of $W$ is directly related to the spectrum of
$M$~\cite{SZWALKS}. Define phases $\varphi_j= \arccos \lambda_j$, so
that $ X^\dagger Y \sket{\phi_j} = \cos \varphi_j \sket{\phi_j}$ The
spectral gap of $M$ is $\delta = 1-\lambda_1 \leq (\varphi_1)^2/2$.
From Eq.~(\ref{uxy}),
\begin{align}
  P_1 U_X^\dagger U_Y^{\;} \sket{\mathfrak 0}\sket{\phi_j} &= \cos{\varphi_j} \sket{\phi_j}\sket{ \mathfrak 0} \\
  P_2 \sket{\phi_j}\sket{\mathfrak 0} &= \cos{\varphi_j}\, U_X^\dagger U_Y^{\;} \sket{\mathfrak 0}\sket{\phi_j} \;,
\end{align}
so $W$ preserves the (at most) two-dimensional subspace spanned by
$\{\sket{\phi_j}\sket{ \mathfrak{0}}, U_X^\dagger U_Y^{\;}
\sket{\mathfrak{0}}\sket{\phi_j}\}$. In terms of the Bloch sphere
defined by states in this subspace, for $j\geq 1$, $W$ acts as a
$4\varphi_j$ rotation along an axis perpendicular to the Bloch-sphere
directions spanned by the defining states~\cite{KOS07}. Thus, the
eigenphases of $W$ in this subspace are $\pm 2 \varphi_j$. The
eigenphase-$0$ states are either the \emph{quantum stationary state}
$\sket{\psi_0} = \sket{\phi_0}\sket{\mathfrak{0}}$ or orthogonal to
both $P_i$. The goal is to prepare $\sket{\psi_0}$ so that we can
sample from the stationary distribution of $M$ by measuring the first
system. (The preparation of $ \sket{\phi_0}$ and its relation
to statistical zero knowledge was studied in~\cite{AhTa2003}.)

To compare a quantum algorithm based on uses of $W$ to the classical
Markov chain algorithm, note that $W$ is readily implemented in terms
of four quantum steps, each of whose complexity is closely related to 
the steps of the classical Markov chain, given our assumptions.
For the purpose of asymptotic comparison, it therefore suffices
to consider the number of quantum steps $W$ versus the number of
classical steps based on $M$.

%%%%%%%%%%%%%%%%%%%%%%%%%%%%%%%%%%%%%%%%%%%%%%%%%%%%%%%%%%%%%%%%%%%%%%%

\noindent {\bf Quantum Simulated Annealing.} We assume that for any
$\beta\geq 0$, there is a transition matrix $M_\beta$ satisfying the
assumptions of the previous section and with stationary distribution
$\pi_\beta^{\sigma}=e^{-\beta E[\sigma]}/\cz$. Like SA, QSA is based
on an annealing schedule that we choose to consist of equally spaced
inverse temperatures $\beta_k = (k-1)\Delta\beta$ for $k=1,\ldots, Q$.
Let $W_k$ be the quantum walk step operator for $M_{\beta_k}$,
$\sket{\psi_0^k}$ its quantum stationary state (\emph{quantum Gibbs
state} for $\beta_k$) and $\varphi_{1,k}$ its phase gap. The goal of QSA
is to sequentially prepare $\sket{\psi_0^{k+1}}$ from
$\sket{\psi_0^k}$ by means of an approximate projective
measurement onto $\sket{\psi_0^{k+1}}$~\cite{CDF02} realized by a
simulated measurement onto the eigenbasis of $W_{k+1}$. We assume that the
uniform superposition $\sket{\psi_0^1}$ can be prepared efficiently.
If the states $\sket{\psi_0^k}$ change slowly enough, the state
$\sket{\psi_0^{Q}}$ can be obtained with high probability of success,
due to a version of the quantum Zeno effect. If $\beta_Q$ is
sufficiently large, $\sket{\psi_0^Q}$ is a good approximation of a
uniform superposition of the ground configurations of $\cs$, so that
we can obtain such a ground configuration with high probability by
measurement. The complexity of QSA is dominated by the complexity of
the simulated measurements, for which we give two strategies, one
based on the phase estimation algorithm (PEA) and the other on
randomized applications of $W_k$. Both strategies' complexities are
dominated by $1/\varphi_{1,k}$. The quadratic quantum speedup is due to
the quadratic increase of $\varphi_{1,k}$ over the eigenvalue gap of
$M_k$.

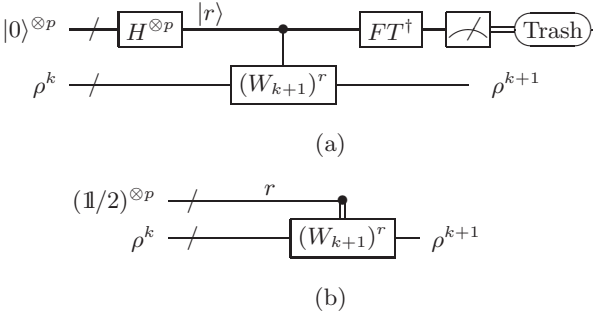
\begin{figure}[ht]
\begin{align*}
\begin{array}{c}  
\Qcircuit @C=1em @R=0.75em {
  \lstick{{\sket 0^{\otimes p}}} &{/} \qw & \gate{H^{\otimes p}} & \raisebox{+.5cm}{{ \mbox{$\sket r$}}} \qw  &\ctrl{1}   & \gate{FT^\dagger} & \meter & \measureCl{\mbox{Trash}}\\
    \lstick{\rho^k} &{/} \qw & \qw &\qw  &\gate{(W_{k+1})^r} &\qw &\qw& \lstick{\rho^{k+1}}
}\\  \\ \text{(a)} \\ \\
  \Qcircuit @C=1em @R=0.75em {
    \lstick{{(\one/2)^{\otimes p}}} &{/} \qw & \qw & \qw &\raisebox{+.3cm}{{ \mbox{$r$}}} \qw  & \control \qw \cwx[1] & & \\
    \lstick{\rho^{k}} &{/} \qw & \qw &\qw &\qw &\gate{(W_{k+1})^r} &\qw & & &\lstick{\rho^{k+1}}
}  \\ \\ \text{(b)}
\end{array}
\end{align*}
\caption{(a) Phase estimation algorithm. The top $p$-qubit register
encodes a $p$-bit approximation to an eigenphase of $W_{k+1}$ on
readout. The second register's states are in $\ch_A \otimes \ch_B$.
The first register is initialized with Hadamard gates to an equal
superposition state in the computational basis. A sequence of $2^p-1$
controlled $W_{k+1}$ operations is applied, and the first register is
measured after an inverse quantum Fourier transform. If the
measurement outcome is $\sket{0}^{\otimes p}$, the second register is
approximately projected onto a $0$-phase eigenstate of $W_{k+1}$. (b)
Randomization procedure. If the PEA's outcome is ignored, the overall
effect on $\ch_A \otimes \ch_B$ is equivalent to the one induced by
initializing a set of $p$ bits (first register) in a random state $r$,
with $r\in\{0,\cdots,2^p-1\}$, and by acting on $\ch_A \otimes \ch_B$
with $(W_{k+1})^r$. Here, double vertical lines indicate classical
control.}
\label{peafig}
\end{figure}

The use of PEA in QSA is depicted in Fig.~\ref{peafig}(a). QSA does
not need to use the result of the phase estimation, though the result
could be used to terminate and restart the procedure if the measurement
outcome is not $\sket{0}^{\otimes p}$. The decoherence it induces in
the eigenbasis of $W_{k+1}$ suffices to achieve the required Zeno effect.
Thus, the effect of the PEA on $\ch_A \otimes \ch_B$ is equivalent to
the one obtained by the action of $r$ $W_{k+1}$'s, with $r$ chosen
uniformly at random from $0$ to $2^p-1$ (Fig.~\ref{peafig}(b)). To
exponentially reduce the error due to remaining coherences between
$\sket{\psi_0^{k+1}}$ and orthogonal states, we repeat the random process
$s$ times, resulting in a total action of $W_{k+1}^{\sum_{q=1}^s r_q}$ with
$0\leq r_q\leq 2^p-1$ independently random. To prevent excessive
amplitude leakage into undesirable $0$-eigenphase eigenstates of
$W_k$, we decohere the second register after each randomization step.
That is, we measure $\ch_B$ in the computational basis and discard the result.
The total complexity of QSA is given by $\co(Q 2^p s)$ walk
steps, where $Q$, $p$ and $s$ are chosen to ensure sufficiently high
probability of success.

Let $\rho^k$ denote the state after the $k$'th randomization and
decoherence step. We have $\rho^{1} = \sket{\psi^{1}_0}\sbra{\psi^{1}_0}$.
Assume that $|\inner{\psi^{k+1}_0}{\psi^{k}_0}|^2 \geq 1-\mu^2$ for
all $k$. By expanding to lowest order in $\Delta\beta$, one can verify
that $\mu = O(\Delta\beta E_M)$. We show by induction that for $2^p >
2^3\pi/\sqrt{2\delta}$ and $s \geq 1+ \log_2(2k)/2 = O(\log(k))$,
$\sbra{\psi^k_0}\rho^k\sket{\psi^k_0} \geq 1-2k\mu^2$. Thus, if $\mu^2
< \epsilon/(4Q)$, $\rho^Q$ is the quantum Gibbs state for
$\beta=Q\Delta\beta$ with error probability at most $\epsilon/2$. We
can write $\rho^k = (1-\chi)\sket{\psi^k_0}\sbra{\psi^k_0} +
\nu_k(\sket{\psi^k_0}\sbra{\psi_\perp^k} + \textrm{H.c.}) +
\chi \rho_{\perp}$, where $\sket{\psi_\perp^k}$ is a unit state
orthogonal to $\sket{\psi_0^k}$, $\rho_\perp$ is a density
matrix with support orthogonal to $\sket{\psi_0^k}$, and $\chi \le 2 k\mu^2$. To make the
induction argument possible, we add the induction hypothesis $\nu_k <
\mu /2$. The induction hypotheses apply to $\rho^1$ by definition. 
Note that $ \sbra{\psi^{k+1}_0}\rho^{k+1}\sket{\psi^{k+1}_0}= \sbra{\psi^{k+1}_0}\rho^k\sket{\psi^{k+1}_0}$.
 We can
estimate $ \sbra{\psi^{k+1}_0}\rho^k\sket{\psi^{k+1}_0} \geq
(1-\chi) |\inner{\psi^{k+1}_0}{\psi^k_0}|^2 -
2\nu_k|\inner{\psi^{k+1}_0}{\psi^k_0}||\inner{\psi^{k+1}_0}{\psi^k_\perp}|
+ \sbra{\psi^{k+1}_0}\rho_\perp\sket{\psi^{k+1}} \geq (1-2k\mu^2)(1-\mu^2) -
2\nu_k\mu \geq 1-2(k+1)\mu^2. $ This establishes the main induction
hypothesis for $k+1$. Before the randomization step, the density
matrix's transition between $\sket{\psi^{k+1}_0}$ and the orthogonal
subspace can be written in the form
$\nu'(\sket{\psi^{k+1}_0}\sbra{\phi_\perp}+\textrm{H.c.})$ with unit
state $\sket{\phi_\perp}$ orthogonal to $\sket{\psi^{k+1}_0}$ and the
other $0$-eigenphase eigenstates of $W^{k+1}$, because the decoherence
step ensures that the support of $P_1$ is preserved by the operator
$\rho^k$. The estimate on
$\sbra{\psi^{k+1}_0}\rho^k\sket{\psi^{k+1}_0}$ implies that
$\nu'\leq\sqrt{2(k+1)}\mu$ by positivity of $\rho^k$~\cite{comment}. Because
$\sket{\psi^{k+1}_0}$ is stabilized by $W_{k+1}$, the transition is
transformed by randomization to
$\nu''(\sbra{\psi^{k+1}_0}\sket{\phi'_\perp}+\textrm{H.c})$ with
$\nu''\sket{\phi'_\perp} = \nu'\left(\frac{1}{2^p}\sum_{r=0}^{2^p-1}
W_{k+1}^{r}\right)^s\sket{\phi_\perp}$. In the eigenbasis of
$W_{k+1}$, the entries of $\sket{\phi_\perp}$ are multiplied by terms
with absolute values $\left(\frac{1}{2^{p}} \left|\sum_{r=0}^{2^p-1}
e^{i r2 \varphi}\right|\right)^s \le \left(\frac 1 {2^{p-1} |1-e^{i
2\varphi}|}\right)^s < \left(\frac {\pi} {2^{p-3} |\varphi|}\right)^s<
2^{-s}$, since the relevant eigenphases $2\varphi$ satisfy
$\pi/2\geq|\varphi|\geq\sqrt{2\delta}$. Thus, the choice
$s=1+ \log_2(2(k+1))/2$ ensures that $\nu''<\mu/2$. Because the decoherence
step preserves $\sket{\psi^{k+1}_0}$, we have
$\nu_{k+1}\leq\nu''<\mu/2$. This completes the induction step of the
proof.

To determine the order of the number of quantum steps $\cn_{QSA}$
required by QSA, let $\beta_f$ be the desired final inverse
temperature, so that $\Delta\beta = \beta_f/Q$. Choose $Q$ to be a
sufficiently large multiple of $\beta_f^2 E_m^2/\epsilon$. For
optimization, we let $\beta_f=\ln(d/(2\epsilon))/\gamma =
\co(\log(d/\epsilon)/\gamma)$. According to the bounds at the
beginning of the previous paragraph, this ensures that after measuring
the final state, the probability of finding a non-optimal configuration is at
most $\epsilon$, with a contribution of $\epsilon/2$ from the
probability of being orthogonal to $\sket{\psi^Q_0}$ and $\epsilon/2$
from the Gibbs distribution's probability of not being optimal.
Because $2^p = \co(1/\sqrt{\delta})$ and $s=O(\log(Q))$, we find that
$\cn_{QSA} = \co(Q \log(Q)/\sqrt{\delta})$ with $Q=O(\beta_f^2
E_m^2/\epsilon)$ and $\beta_f = \co(\log(d/\epsilon)/\gamma)$. If we
anticipate that $Q>d$, we can just search every configuration classically to
find the optima, so we can bound $\log(Q)\leq\log(d)$ to simplify
\begin{equation}
\cn_{QSA} =\co\( \(\frac{E_M}{\gamma}\)^2\frac{\log^2(d /\epsilon)\log d} {\epsilon \sqrt{\delta}} \)\;. 
\label{nqsa}
\end{equation}
The dependence of $\cn_{QSA}$ on $1/\epsilon$ can be improved to
$\textrm{polylog}(1/\epsilon)$ by repetition of QSA with an initial
target error $\epsilon=1/2$ in Eq.~(\ref{nqsa}). For optimization, it
suffices to repeat QSA $O(\log(\epsilon))$ many times. Another
approach that may be used to prepare the desired stationary state with
high probability of success is to apply a high-confidence
version of the PEA~\cite{KOS07} at the end of QSA to project onto
$\sket{\psi_0^Q}$, the stationary state for inverse temperature
$\beta_f$. If the projection fails, the algorithm is repeated.

Although the dependence of $\cn_{QSA}$ on $E_M/\gamma$ is worse than
the one appearing in classical SA, it is worth noting that unlike the
inverse spectral gap $1/\delta$, in many important applications this
parameter is bounded by a constant or a polynomial in instance size.

%%%%%%%%%%%%%%%%%%%%%%%%%%%%%%%%%%%%%%%%%%%%%%%%%%%%%%%%%%%%%%%%%%%%%%%
\noindent {\bf Conclusions.} We presented a quantum algorithm
based on a ``quantization'' of simulated annealing algorithms
implemented with MCMC methods. This quantum simulated annealing (QSA)
algorithm forces the state to closely follow a superposition with
amplitudes derived from finite-temperature Gibbs distributions. This
is accomplished by either an explicit measurement using phase
estimation with quantum walk operators, or by decoherence using random
applications of these operators. QSA can be used both for
combinatorial optimization and for sampling from a Gibbs distribution
for statistical physics applications. In contrast to SA, which scales
with $\cO(1/\delta)$, where $\delta$ is the minimal spectral gap of
the transition matrices, QSA scales with $\cO(1/\sqrt \delta)$.
Although in general the QSA does not yield a polynomial-resource
algorithm, it reduces required resources by an asymptotic exponential
factor for the ubiquitous hard cases, where the gap becomes
exponentially small in the problem size.

\acknowledgments We thank Stephen Jordan for discussions, and Bryan
Eastin and David Hume for their careful reading of the manuscript.
This research was supported by Perimeter Institute for Theoretical
Physics, by the Government of Canada through Industry Canada and by
the Province of Ontario through the Ministry of Research and
Innovation. This work was partially carried out under the auspices of
the NNSA of the US DOE at LANL under Contract No. DE-AC52-06NA25396,
and with support from NSF Grant No. PHY-0653596. Contributions to this
work by NIST, an agency of the US government, are not subject to
copyright laws.

%%%%%%%%%%%%%%%%%%%%%%%%%%%%%%%%%%%%%%%%
%%%%%%%%%%%%%%%%%%%%%%%%%%%%%%%%%%%%%%%%


\begin{thebibliography}{10}
\bibitem{comment}
For a density matrix $\rho$, we have $|\rho_{lm}| \le \sqrt{\rho_{ll} \rho_{mm}}$,
with $\rho_{ll'}=\bra{l} \rho \ket{l'}$.
\end{thebibliography}
\end{document}